\documentclass[conference]{IEEEtran}
\usepackage{url}
\usepackage{cite}
\usepackage{amsmath,amssymb,amsfonts}
\usepackage{algorithmic}
\usepackage{graphicx}
\usepackage{textcomp}
\usepackage{xcolor}
\usepackage{tabularx} 
\begin{document}
\title{A Survey of Security, Privacy, and Ethics of Large Language Models in Telecom}

\author{\IEEEauthorblockN{Joy Bose}
\IEEEauthorblockA{\textit{Ericsson}\\
Bangalore, India \\
joy.bose@ericsson.com}
\and
\IEEEauthorblockN{Srinivas Adyapak}
\IEEEauthorblockA{\textit{Ericsson}\\
Bangalore, India \\
srinivas.adyapak@ericsson.com}
}

\maketitle

\begin{abstract}
Large Language Models (LLMs) are increasingly integrated into telecommunications for customer service, network analytics, and operational support. While offering efficiency and innovation, their deployment raises critical privacy, security, and ethical challenges due to the uniquely sensitive and longitudinal nature of telecom data. This paper presents a survey of LLM risks and safeguards in telecom context. Our survey introduces a telecom-specific taxonomy linking privacy, security, and ethics with use cases and outlines open research challenges such as longitudinal data privacy, cross-border compliance, and scalable guardrails.
\end{abstract}


\section{Introduction}
Recent initiatives such as the Global Telco AI Alliance, TM Forum’s Open Digital Architecture (ODA) \cite{oda2023guide, gsma2024toolkit} and GSMA’s AI for Impact Toolkit illustrate a telecom industry-wide move toward secure, regulation-compliant LLM systems. This paper surveys security, privacy, and ethical dimensions of LLM deployments in telecom, focusing on domain-specific threats, mitigations, and research directions.

\section{Privacy and Security Issues in Telecom Data}
Telecom networks generate and manage vast volumes of sensitive and personal information. Telecom datasets include not only subscriber identity and contact details, but also real-time metadata such as call logs, SMS history, geolocation traces, internet browsing activity, and device information. When integrated into Large Language Model (LLM) workflows, such data becomes particularly vulnerable due to the black-box nature and general-purpose behavior of LLMs.

Telecom data is uniquely risky compared to other domains. Unlike healthcare data, telecom data is continuous and behaviorally complete. Unlike finance, telecom logs capture movement and communication metadata simultaneously. This combination increases re-identification risk and makes anonymization brittle over time.

\subsection{Types of Sensitive Data Involved}
The scope and sensitivity of telecom data present unique risks due to sensitive data, including Personally Identifiable Information (PII) such as names and addresses, communication metadata such as call detail records (CDRs) and network access logs, location data such as triangulated cell tower data, GPS traces, and service interaction logs of customer queries, chatbot conversations, and support tickets used for LLM fine-tuning. Leakage of such information via LLM output or training logs can result in significant damage.

\subsection{Key Privacy Challenges with LLMs}
Several characteristics of telecom LLMs complicate traditional privacy protections, including memorization risk where LLMs can memorize and regurgitate personal data inappropriately, and lack of consent, where users whose conversations are used for training may not have explicitly consented to such usage of their data. 

\subsection{Security Vulnerabilities Specific to Telecom LLMs}

Telecom infrastructure is already a high-value target for cyberattacks, and the introduction of LLMs introduces new attack surfaces:
\begin{itemize}
    \item \textbf{Prompt Injection}: Malicious inputs  by users (e.g. "Ignore prior instructions and ...") can hijack the LLM’s behavior, potentially revealing sensitive backend data or bypassing restrictions \cite{greshake2023promptinjection, owasp2023llm, owasp2025genai}.
    \item \textbf{Model Inversion Attacks}: Adversaries may extract training data by observing outputs in response to cleverly constructed prompts. \cite{fredrikson2015inversion}
    \item \textbf{Membership Inference Attacks}: Attackers can determine whether specific records were part of the LLM’s training set \cite{shokri2017membership}.
    \item \textbf{Model Stealing and Cloning}: Exposed LLM endpoints may be probed to replicate proprietary models and architectures \cite{shayegani2023llm, das2024survey}.
    \item \textbf{Plugin Exploits}: Third-party integrations and plugin systems often lack robust sandboxing, posing additional risks.
\end{itemize}

Real-time, high-volume deployment across RAN, Core, and OSS/BSS planes amplifies these vulnerabilities and increases the risk of unintended data exposure.

Industry practitioners note that telecom networks operate across RAN, Core, and OSS/BSS planes, each introducing unique attack surfaces \cite{shorov2024}.

Autonomous LLM agents used for applications like workflow automation or triaging can introduce new risks. When agents call tools or communicate with other agents, the attack surface expands to agent states, memory stores, and API interfaces. A compromised agent can escalate access, trigger unintended actions, or spread malicious instructions across chained agents. Because they operate independently, detecting misuse or unexpected access is harder—especially in cloud-hosted environments. Access to agent memory is effectively access to the network, making strict controls and monitoring essential.

\section{Related Work in Privacy Aspects}
Recent research and industry practices have focused significantly on improving privacy preservation for Large Language Models (LLMs), especially in domains like telecom where data sensitivity is high. 

Key privacy enhancing techniques include Differential Privacy (DP), Federated Learning (FL) and Red Teaming (RT). DP introduces carefully calibrated noise during model training to prevent leakage of specific data entries while preserving overall model utility. \cite{priyanshu2024usenix}. In telecom contexts, integrating DP can prevent models from memorizing customer conversations, authentication records, or contact logs. 
FL allows models to be trained across decentralized data sources. In telecom, FL has potential applications in distributed anomaly detection, user behavior prediction, or personalization without compromising privacy (\cite{mcmahan2017deepnetworks, thakkar2020memorization} ).
RT focuses on proactively identifying potential failure modes in LLMs by simulating adversarial behavior and attack patterns. In telecom deployments, RT can identify vulnerabilities such as data echoing, uncontrolled generation, and prompt hijacking, especially in customer service interfaces. \cite{das2024survey, shayegani2023llm}

Data minimization principles suggest that only essential and anonymized data should be used for model training or inference. 

Basic but essential techniques like removing names, account numbers, and other identifiers are still widely used to protect privacy. These systems, when combined with knowledge retrieval (RAG) \cite{lewis2020rag, alabbasi2025teleoracle} and metadata tagging, can ensure safe data flow through the model stack.

Table \ref{tab:ai_security} summarizes some standard privacy enhancing techniques and how they relate to telecom. 

\begin{table}[h]
\centering
\caption{Privacy Enhancing Techniques in Telecom}
\begin{tabular}{|p{1.4cm}|p{3.4cm}|p{2.0cm}|}
\hline
\textbf{Technique} & \textbf{Benefit} & \textbf{Limitation} \\ \hline
DP & Prevents CDR memorization & Accuracy drop \\ \hline
FL & Training across edge nodes & High comm.\ cost \\ \hline
RT & Detects prompt leaks & No quant.\ metric \\ \hline
\end{tabular}
\label{tab:ai_security}
\end{table}

Despite growing attention, measuring privacy in LLMs remains a non-trivial challenge. Privacy attacks may succeed only under rare or complex conditions, making it difficult to reproduce them consistently. Moreover, telecom companies may be reluctant to share training datasets or breach reports due to reputational concerns, limiting research transparency.

There is a growing call for open, community-validated datasets and auditing tools that allow consistent privacy evaluations of telecom-oriented LLMs (\cite{google2023privacy, privacyinternational2023, fpf2024} ). Such frameworks would allow telcos to compare models and validate claims before deployment.

\section{Related Work in Security Aspects}

As LLMs are integrated into telecommunications infrastructure, the threat landscape becomes significantly more complex. In addition to traditional IT and network vulnerabilities, telecom operators must now account for the unique security challenges posed by foundation models. This section surveys the body of research and practice focused on fortifying LLMs against various security threats in telecom use cases.

In industry analyses, AI/ML in telecom is viewed in three concurrent roles—as tools leveraged by attackers, as defensive enablers, and as attack targets themselves \cite{shorov2024}.

\subsection{Adversarial Prompting and Jailbreaking}

One of the most prominent LLM-specific attack vectors is adversarial prompting, also known as jailbreaking. Here, malicious users try inputs to bypass content filters, cause models to behave erratically, or reveal restricted information \cite{greshake2023promptinjection, wu2023jailbreak, wiz2024}. For instance, attackers may ask queries that produce internal configuration data, network topologies, or customer records.

One strategy is to use LLMs in telecom-specific cyber defense, such as intrusion detection in satellite and vehicle networks, which can enhance defenses against adversarial prompting by leveraging domain-specific knowledge \cite{hassanin2024review}. Researchers have also proposed a systematic generation of adversarial suffixes through greedy and gradient-based optimization, demonstrating universal and transferable attacks across both open-source and proprietary LLMs like GPT and Claude. Such attacks can induce models to output objectionable content despite fine-tuning and safety layers \cite{zou2023adversarial}.

\subsection{Model Inversion and Membership Inference Attacks}

Model inversion attacks attempt to reconstruct input data by observing a model’s outputs. This poses serious risks in telecom, where leaked outputs may expose personally identifiable or proprietary data \cite{fredrikson2015inversion}. 

Membership inference attacks, on the other hand, allow adversaries to determine whether a specific sample was included in the model’s training dataset \cite{shokri2017membership}. This is especially problematic if LLMs are fine-tuned on customer interactions or usage logs without strong anonymization or aggregation protocols. Techniques to reduce this risk include dropout-based regularization, entropy smoothing, and differential privacy.

\subsection{Use of Guardrails and Policy Layers}

To improve runtime safety, industry tools like Nvidia NeMo Guardrails \cite{nemo2023guardrails} and OpenAI’s moderation API allow developers to define constraints on model behavior. NeMo Guardrails is a framework that lets developers define conversation rules, safety policies, and response validation logic for LLM-powered applications. It has programmable components such as policy servers, re-ranking engines, and content classifiers to enforce safety in real-time. Telecom LLMs can benefit from NeMo Guardrails by defining domain-specific filters such as for customer service and restricting access to confidential operational commands. Such rule engines and policy enforcement layers allow telcos to control what kind of answers can be generated in regulated contexts.

\subsection{Research Gaps and Future Challenges}
Despite active work in security, some gaps remain such as the following:
\begin{itemize}
    \item Most LLM security research is general-purpose and not tailored to telecom’s real-time and regulated environment.
    \item Integration of classical network security tools with LLM-specific defenses remains primitive.
    \item Proactive security-by-design principles are often missing in early-stage telecom AI projects.
    \item Trade-offs between performance, cost, and robust security remain poorly understood.
\end{itemize}
Efforts like AdvBench and MITRE ATLAS \cite{mitre2023atlas} highlight the need for domain-specific testbeds to evaluate adversarial robustness, including LLM responses to spoofed tickets or intrusion attempts.
Cloud-hosted telecom LLMs add risks from IAM misconfigurations, multi-tenant isolation failures, and exposed agent logs. Anyone with access to embeddings, vector stores, or agent histories can leak telecom data. Misconfigured API gateways or CI/CD pipelines further widen the attack surface. Separation of duties, encrypted memory, and strong audit trails are essential in hybrid cloud telecom deployments.

\section{Ethical Considerations}
Ethical concerns \cite{ericsson2022ethicalaitelecom, jiao2024llmethics} encompass a broad set of issues, including bias, transparency and governance. 

From an operational perspective, ethical risks in telecom LLMs arise when longitudinal KPIs are summarized without fixed cohort definitions, explicit event calendars, or disclosure of missing intervals, leading to overconfident or biased conclusions. Embedding constraints that prohibit causal claims, require stratified reporting, and surface uncertainty helps align LLM outputs with responsible decision-making in network and service management.

Algorithmic bias is where LLMs may exhibit skewed behavior that unfairly disadvantages certain user groups and prioritize others \cite{bolukbasi2016nips}, such as selectively escalating cases related to telecom billing disputes. Mitigation techniques for bias include debiasing during pretraining through dataset balancing, post-hoc calibration and filtering of biased outputs and regular fairness audits. LLMs are typically black-box systems, making it difficult to understand why a specific response was generated. Approaches to improve transparency \cite{ribeiro2016trust} include generating rationales or "thought traces" along with model outputs, integrating interpretable sub-models like decision trees in hybrid architectures and logging the sources behind a model’s decision. Also, telecom operators must be accountable for the AI systems they deploy. Governance frameworks such as the EU AI Act, IEEE Ethically Aligned Design, and ISO/IEC 42001 provide structured guidance for aligning LLM usage with societal expectations \cite{euaiact2024, ieee2019ethical}. 

\begin{figure}[h]
\centering
\includegraphics[width=8cm]{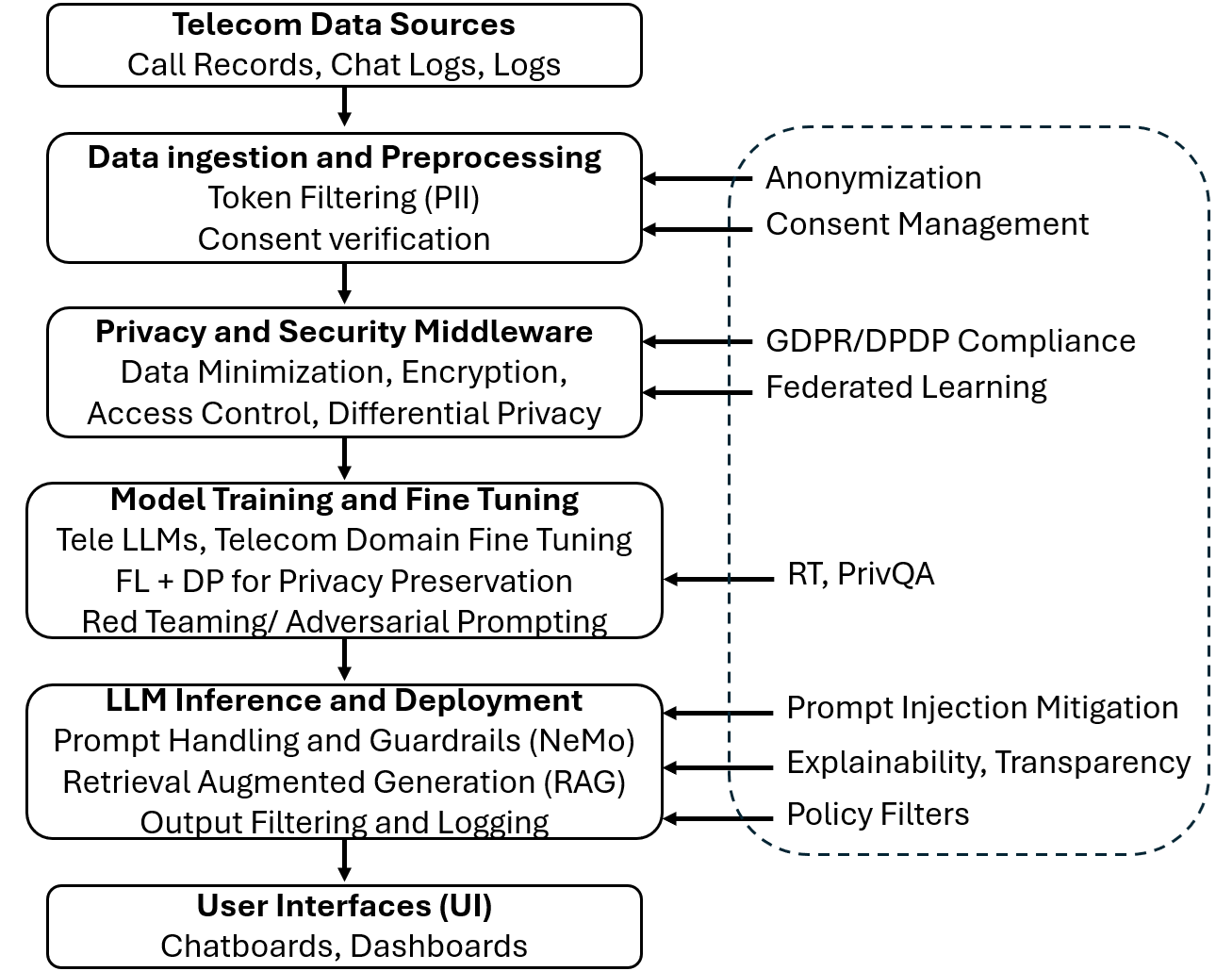}
\caption{Architecture of telecom LLM deployment highlighting stages vulnerable to privacy, security, and ethical issues, along with mitigation strategies such as differential privacy, federated learning, and guardrails}
\label{fig:telecomllm}
\end{figure}

\begin{table*}[t]
\centering
\caption{Mapping of telecom data types to LLM risks, countermeasures, and typical use cases}
\label{tab:telecom_llm_taxonomy}
\begin{tabularx}{\linewidth}{|p{3cm}|X|X|X|}
\hline
\textbf{Telecom Data Type} & \textbf{Primary Risk in LLM Context} & \textbf{Representative Countermeasure} & \textbf{Example Telecom Use Case} \\ \hline

Call Detail Records (CDRs) &
Memorization of subscriber metadata and movement patterns &
Differential Privacy (noise injection during training) &
LLM-based customer analytics or fraud detection \\ \hline

Subscriber Profiles / KYC Data &
Re-identification and prompt leakage of personally identifiable information (PII) &
Federated Learning with secure aggregation &
LLM-driven recommendation engine for plans \\ \hline

Network Logs (OSS/BSS) &
Prompt injection or adversarial queries exposing configurations &
Guardrail policies, content filters, and prompt moderation &
Chatbot for network fault diagnosis \\ \hline

Location and Usage Telemetry &
Cross-border data transfer and regulatory non-compliance &
Privacy-enhancing technologies (PETs) with geo-fencing &
Predictive maintenance or mobility forecasting \\ \hline

Customer Interactions (Chats, Tickets) &
Hallucinated responses or bias in support automation &
Human-in-loop review with explainability logging &
Automated customer service LLMs \\ \hline

Billing and Payment Records &
Algorithmic bias in charging logic (ethical risk) &
Transparent audit trails and DPDP compliance &
LLM-assisted billing dispute resolution \\ \hline

\end{tabularx}
\end{table*}

\section{Emerging Directions}
Figure~\ref{fig:telecomllm} shows the architecture of an LLM deployment showing stages with security and privacy issues and their remedies.
Several forward-looking research and industry initiatives are addressing the challenges outlined above. Current research and development trends include the following:
\begin{itemize}
\item{Tele-LLMs for domain-specific fine-tuning}: Domain-specific models trained on 3GPP specs, call center transcripts, and OSS/BSS logs show better performance in telco contexts \cite{maatouk2024telellm}.

\item{RAG with Guardrails Architectures}: Retrieval-Augmented Generation combined with filtering layers is gaining traction for controlled, dynamic outputs \cite{lewis2020rag, alabbasi2025teleoracle}. 

\item{PET (Privacy Enhancing Technologies) Combinations}: New architectures are blending Federated Learning (FL), Differential Privacy (DP), and Secure multi-party computation (SMPC) to create more robust privacy guarantees \cite{abadi2018deeplearning, mcmahan2017deepnetworks, ryffel2019nips}.

\item{Secure LLMOps pipelines}: MLsecOps pipelines are being adopted to manage model lifecycle securely, from fine-tuning to decommissioning. LLMOps frameworks mirrors the development of MLOps, providing tools to manage training, deployment, versioning, and monitoring of LLMs in production. LLMOps tools should include model registries, lineage graphs, and usage monitoring to detect drift, toxic behaviors, or emerging vulnerabilities. 

\end{itemize}

Additionally, industry practitioners recommend standardized secure AI/ML protocols, continuous risk assessment updates, and the promotion of privacy-enhancing technologies as priority directions for the telecom sector \cite{shorov2024}.

\section{Analysis}
In synthesizing the current state of literature and practice, several important themes and observations emerge regarding the deployment of LLMs in the telecommunications sector.

\subsection{Performance vs. Privacy Trade-offs}
One of the most consistent trade-offs identified is between model performance and privacy preservation. Techniques like differential privacy and data minimization are effective in reducing the leakage of sensitive data, but they often come at the cost of reduced model accuracy or fluency. For example, LLMs trained with differential privacy may generalize less effectively, especially on nuanced or technical telecom queries. Telecom operators must therefore carefully assess use cases and choose privacy-preserving methods that meet regulatory requirements without severely compromising utility and performance.

Autonomous network agents increasingly assist with fault triage, log summarization, and OSS/BSS queries. If one agent is compromised, it can direct other agents to pull sensitive logs or modify workflows, causing cascading failures. Preventing this requires lineage tracking of agent actions and access-controlled memory.

\subsection{Security Mechanisms are Still Immature}

Despite growing awareness of LLM-specific vulnerabilities, industry-wide security practices are still evolving. Many telecom firms rely on third-party models accessed via APIs (such as GPT-4 or Claude), which exposes them to external threats, potential model inversion, and supply chain risks \cite{shayegani2023llm, ranaldi2022pretrained, mo2020darknetz}. Even internally hosted models may be susceptible to prompt injection and adversarial attacks if not adequately sandboxed and monitored. While frameworks like red teaming and secure DevOps pipelines are promising, they are not yet consistently adopted across the telecom sector.

\subsection{Ethics, regulation and oversight gaps}
Ethical dimensions like explainability, fairness, and consent are often overlooked in telecom LLM deployments, leading to a gap between policy mandates and real-world practice. Regulatory compliance is fragmented across regions (such as GDPR vs DPDP), requiring modular AI architectures for lawful deployment. Industry maturity also varies, where larger telcos build proprietary systems, while smaller players rely on third-party models with limited governance. 

In addition, Human oversight and hybrid systems, such as combining LLMs with rule-based systems or traditional NLP tools, are essential to mitigate automation risks. A multi-dimensional risk classification for telco systems can aid in governance. For example, a single telecom LLM application such as a network fault assistant might rate Level 5 in security, but only Level 2 in privacy if consent is unclear. 

\subsection{Emerging Best Practices}
Some emerging best practices include: Least-privileging IAM for LLM endpoints, embeddings, and agent memory, having an action-validation layer before any agent triggers tools/APIs, tracking agent lineage to detect cascading misuse, encrypting vector stores and enforce memory access policies, continuous policy audits and anomaly detection on agent actions and segmenting cloud environments and rotate secrets aggressively.

\section{Conclusion}
LLMs are poised to transform telecom, enabling new efficiencies in customer service, network analytics, and operational support. Yet their deployment also exposes the sector to heightened privacy, security, and ethical risks due to the persistent, longitudinal, and cross-border nature of telecom data. This survey contributes to addressing these challenges in a few ways. First, we ensured coverage of both academic and industry work while focusing on telecom-specific applications. Second, we conducted a comparative analysis of privacy and security enhancing techniques contextualized with telecom use cases. Finally, we formulated a set of open research challenges that frame future work in areas such as longitudinal data privacy, cross-border compliance, scalable red-teaming, and the development of domain-specific benchmarks. As LLMs continue to permeate telecom operations, progress will require a multidisciplinary effort that embeds trust, accountability, and regulatory alignment into every layer of the LLM lifecycle. Secure telecom LLMs will require  progress in privacy-enhancing design, robust guardrails, and cross-border compliance benchmarks grounded in real network data.

\end{document}